\documentclass[runningheads]{llncs}
\usepackage[T1]{fontenc}
\usepackage[protrusion=true,expansion=false]{microtype}
\usepackage{graphicx}
\usepackage{graphicx,verbatim}
\usepackage[utf8]{inputenc}
\usepackage{pifont}
\usepackage{tikz}
\usetikzlibrary{shapes, arrows.meta, positioning, calc}
\usepackage{amsmath, amssymb, amsfonts}    % <--- CORRECT
\usepackage{mathtools}
\usepackage{euscript}
\usetikzlibrary{shadows}
\usepackage{graphicx}
\graphicspath{{Figures/}} % Note the double brackets usually required for path safety
\usepackage{subcaption}
\usepackage{placeins}
\usepackage{caption} % Note: llncs sometimes warns about this, but it's needed for subcaption
\usepackage{float}
\usepackage[table]{xcolor} 
\usepackage{booktabs}
\usepackage{multirow}
\usepackage{adjustbox}
\usepackage{tabularx, array}
\usepackage{colortbl}
\usepackage{float}
\usepackage[table]{xcolor}
\usepackage{adjustbox}
\definecolor{bestgreen}{RGB}{200,230,201}
\definecolor{secondyellow}{RGB}{255,245,200}
\newcommand{\best}[1]{\cellcolor{bestgreen}#1}
\newcommand{\sbest}[1]{\cellcolor{secondyellow}#1}
\usepackage{algorithm}
\usepackage{algpseudocode} % Kept this, removed 'algorithmic'
\usepackage[most]{tcolorbox} 
\usepackage{verbatim}
\usepackage{comment}
\usepackage[hidelinks]{hyperref}
\usepackage[capitalise,noabbrev]{cleveref}
\hypersetup{
    colorlinks=true,
    linkcolor=blue,
    filecolor=cyan,      
    urlcolor=magenta,
}
\begin{document}
\title{Echo-\underline{E\textsuperscript{3}}Net: \underline{E}fficient
\underline{E}ndocardial Spatio-Temporal Network for \underline{E}jection Fraction Estimation}
\titlerunning{Echo-E\textsuperscript{3}Net}

\author{
Moein Heidari\inst{1}\textsuperscript{*} \and
Afshin Bozorgpour\inst{2}\textsuperscript{*} \and
AmirHossein Zarif-Fakharnia\inst{3} \and
Wenjin Chen\inst{4} \and
Dorit Merhof\inst{2} \and
D. J. Foran\inst{4} \and
Jasmine Grewal\inst{1} \and
Ilker Hacihaliloglu\inst{1}
}

\authorrunning{M. Heidari et al.}

\institute{
University of British Columbia, Vancouver, BC, Canada\\
\email{moein.heidari@ubc.ca}
\and
University of Regensburg, Regensburg, Germany
\and
Iran University of Science and Technology, Tehran, Iran
\and
Rutgers Cancer Institute of New Jersey, New Brunswick, NJ, USA\\[2pt]
\textsuperscript{*}Equal contribution.
}

\maketitle              % typeset the header of the contribution
\begin{abstract}

Left ventricular ejection fraction (LVEF) is a primary marker of cardiac
function. However, routine estimation from endocardial measurements requires manual delineation at end-diastole (ED) and end-systole (ES), a process that is time-consuming and subject to inter-observer variability.
Reliable automation is especially valuable for point-of-care ultrasound
(POCUS), where computational resources are limited and the acquisition quality varies. We propose
Echo-E\textsuperscript{3}Net, an anatomy-guided spatio-temporal network that
explicitly embeds cardiac anatomy into LVEF prediction. A dual-phase
Endocardial Border Detector (E\textsuperscript{2}CBD) uses phase-specific
cross-attention to localize ED/ES endocardial landmarks and produce
phase-aware landmark embeddings, while an Endocardial Feature Aggregator
(E\textsuperscript{2}FA) fuses these embeddings with global statistical
descriptors of deep feature maps to refine EF regression. Training is guided by a lightweight geometric loss that uses ED and ES endocardial landmarks to regularize EF prediction. On EchoNet-Dynamic and a PSAX subset of EchoNet-Pediatric, Echo-E\textsuperscript{3}Net attains competitive performance using only 1.55M parameters and 8.05 GFLOPs, an order-of-magnitude compute reduction
versus recent baselines, supporting real-time deployment. Our code is publicly available at~\href{https://github.com/moeinheidari7829/Echo-E3Net}{\textcolor{magenta}{GitHub}}.

\end{abstract}

\section{Introduction}
\label{sec:intro}

Left ventricular ejection fraction (LVEF) is a key indicator of cardiac
function, routinely used to assess heart failure and guide clinical decision
making~\cite{mokhtari2022echognn,light,muhtaseb2022echocotr,thomas2024echonarrator,lai2024echomen}.
In standard adult echocardiography, LVEF is commonly estimated from end-diastole (ED) and end-systole (ES) endocardial measurements, often using the method of disks when apical views are available~\cite {light}. This
process is time-consuming and subject to substantial inter-observer
variability, particularly when image quality is reduced, or views are
suboptimal~\cite{real_time,maani2024coreecho}. These limitations are
amplified in point-of-care ultrasound (POCUS), which is increasingly used
for bedside cardiac assessment but is highly operator dependent and runs on
hardware with constrained compute~\cite{point_of_care,real_time}. Reliable
LVEF estimation in this setting, therefore, demands methods that are not only
accurate but also computationally efficient and robust enough for real-world
deployment~\cite{heidari2024enhancing,ejec_1}.

Deep learning has substantially advanced automated LVEF
estimation~\cite{muhtaseb2022echocotr,mokhtari2022echognn,lai2024echomen}. Existing
approaches include segmentation-based
pipelines~\cite{ouyang2020video,dai2022cyclical}, direct video-level
regression models~\cite{ouyang2020video,kazemi2020deep,muhtaseb2022echocotr},
graph and keypoint-guided strategies that encode cardiac
structure~\cite{mokhtari2022echognn,thomas2022light}, and attention-based
designs for spatio-temporal representation
learning~\cite{mokhtari2023gemtrans}. Graph-based methods such as
EchoGNN~\cite{mokhtari2022echognn} and EchoGraphs~\cite{thomas2022light}
improve interpretability through keypoint representations and shape
constraints, but can underemphasize the ED and ES frames that are central
to clinical EF measurement. EchoMEN~\cite{lai2024echomen} addresses EF-range
imbalance with a multi-expert design at the cost of added computation, while
CoReEcho~\cite{maani2024coreecho} and reconstruction-based
priors~\cite{yang2025cardiacnet} further improve accuracy. Across these
methods, however, strong accuracy is consistently coupled with heavy
parameter counts and FLOPs, which limits adoption in resource-constrained
POCUS settings~\cite{ejec_1}.

We address this gap with Echo-E\textsuperscript{3}Net, an efficient
endocardial spatio-temporal network that explicitly embeds cardiac anatomy
into LVEF prediction. Specifically, a dual-phase Endocardial Border Detector
(E\textsuperscript{2}CBD) lets ED and ES queries attend to compressed
multi-scale spatio-temporal tokens, yielding explicit landmark coordinates
and phase-aware landmark embeddings that compactly encode endocardial
geometry and motion. An Endocardial Feature Aggregator
(E\textsuperscript{2}FA) fuses these embeddings with global statistical
descriptors of deep feature maps to refine EF regression. Training is guided by a lightweight geometric objective that regularizes ED and ES landmark geometry. Our
contributions are threefold:
\ding{202}\ an anatomy-guided architecture in which dual-phase
cross-attention explicitly localizes ED/ES endocardial landmarks, keeping EF
prediction grounded in cardiac structure;
\ding{203}\ a single-view geometric regularizer that supervises ED and ES landmark geometry during training;
\ding{204}\ competitive accuracy on EchoNet-Dynamic and on a PSAX subset of EchoNet-Pediatric, at an order-of-magnitude
lower compute (1.55M parameters, 8.05 GFLOPs), with real-time CPU inference
and no pre-training, augmentation, or ensembling.

\section{Method}
\label{sec:method}

\begin{figure}[t]  
  \centering
  \includegraphics[width=1\textwidth]{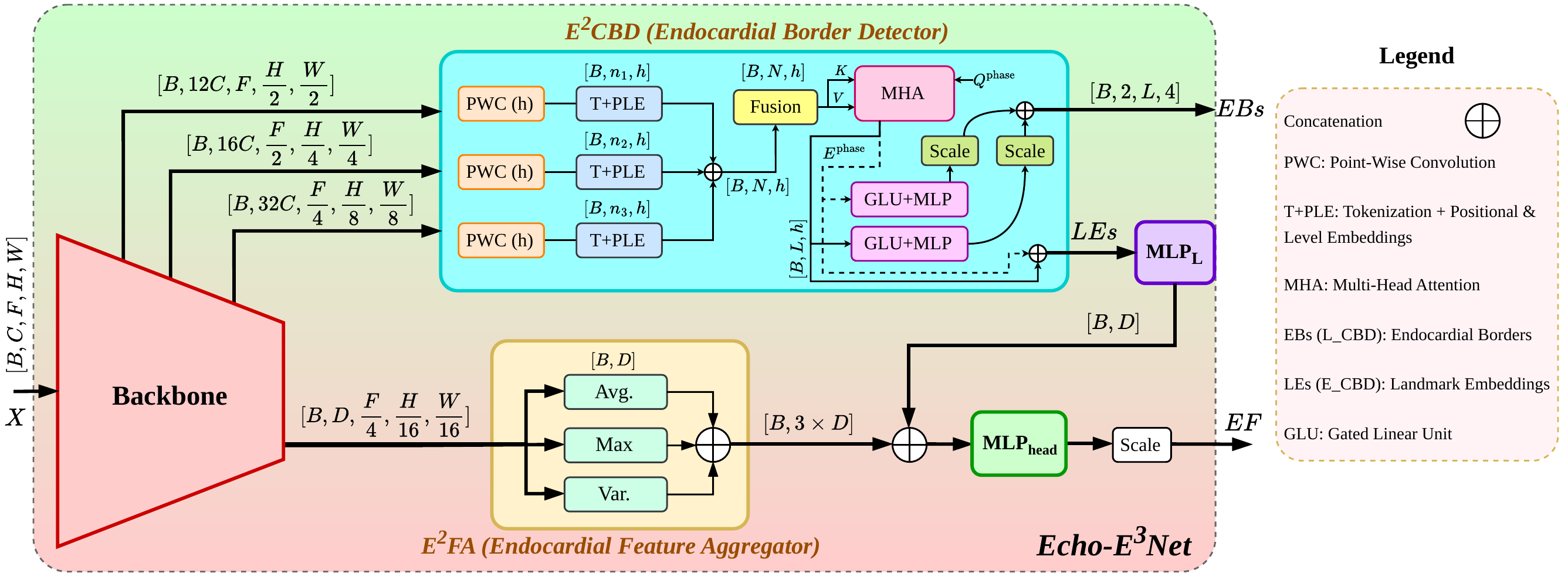}  % Adjust width as needed; filename without quotes
  \caption{Architecture of Echo-E\textsuperscript{3}Net. An LHUNet encoder extracts multi-scale spatio-temporal features from the input video $X$. E\textsuperscript{2}CBD tokenizes the skip features and applies phase-specific cross-attention from ED/ES landmark queries to predict dual-phase endocardial border chords (EBs) and landmark embeddings (LEs). E\textsuperscript{2}FA fuses global feature statistics (average, maximum, variance) with the landmark descriptor to regress EF. Here, EBs denote the predicted endocardial border chords and LEs the landmark embeddings.}
\label{fig:arch}
\end{figure}

Given an echocardiographic video $X \in \mathbb{R}^{B \times C \times F
\times H \times W}$ with reference LVEF $y \in (0,100)$ and dual-phase
(ED/ES) endocardial landmark annotations, our goal is to predict EF while
remaining consistent with the underlying cardiac geometry. As shown in
\Cref{fig:arch}, Echo-E\textsuperscript{3}Net comprises three components:
(i)~a backbone encoder that extracts multi-scale 3D features; (ii)~a
dual-phase Endocardial Border Detector (E\textsuperscript{2}CBD) that predicts
ED/ES landmark chords and phase-aware landmark embeddings; and (iii)~an
Endocardial Feature Aggregator (E\textsuperscript{2}FA) that fuses these
geometric descriptors with global feature statistics to regress EF.
E\textsuperscript{2}CBD is trained with anatomically motivated constraints inspired by disk-based ventricular measurement, so that both local shape and global contractility cues drive the final prediction.

\subsection{Backbone Network}
We adopt the encoder of LHUNet~\cite{sadegheih2025lhu} as a lightweight
volumetric feature extractor. Unlike its original segmentation use, we keep
only the encoder: initial convolutions adapt the input channels, subsequent
blocks extract local features while downsampling, and hybrid blocks with
large-kernel convolutions and spatial/channel attention capture longer-range
context. From $X$, the encoder produces three multi-scale skip features
$\{s_1, s_2, s_3\}$ of decreasing spatial and temporal resolution, together
with a deepest feature map $x$. The
skip features feed E\textsuperscript{2}CBD; the deepest feature feeds
E\textsuperscript{2}FA.

\subsection{Dual-Phase Endocardial Border Detector (E\texorpdfstring{$^2$}{2}CBD)}
Clinical EF estimation relies on endocardial boundaries delineated at ED and ES.
E\textsuperscript{2}CBD mirrors this by predicting, for each phase, a set of
$L$ landmark chords ordered along the ventricular boundary, where each chord is a pair of opposing endocardial points.

\noindent\textbf{Multi-scale tokenization.}
Each skip feature $s_i$ is projected to a shared hidden width $h$ by a
point-wise convolution (PWC), $\tilde{s}_i = \mathrm{PWC}(s_i) \in
\mathbb{R}^{B \times h \times F_i \times H_i \times W_i}$, then flattened
over spatial/temporal axes into $n_i = F_i H_i W_i$ tokens. We add a positional and a level embedding (T+PLE in \Cref{fig:arch}):
$P_{\text{pos}}$ is a learned linear map of a normalized spatio-temporal
coordinate grid $(t,y,x)\!\in\![-1,1]^3$, so that temporal and spatial
positions are encoded jointly, and the learned vector
$p_{\mathrm{lvl},i}\!\in\!\mathbb{R}^h$ identifies the feature scale:
\begin{equation}
f_i = \mathrm{flatten}(\tilde{s}_i) + P_{\text{pos}}(t,y,x) + p_{\mathrm{lvl},i}
\;\in\; \mathbb{R}^{B \times n_i \times h}.
\end{equation}
The three levels are concatenated into a multi-scale token set of
$N=\sum_i n_i$ tokens and refined with a lightweight token-wise MLP (Fusion):
\begin{equation}
z = \mathrm{Fusion}\big([\,f_1;f_2;f_3\,]\big) \in \mathbb{R}^{B \times N \times h}.
\end{equation}

\noindent\textbf{Phase-specific cross-attention and decoding.}
Two learnable query banks $Q^{\mathrm{ED}}, Q^{\mathrm{ES}} \in
\mathbb{R}^{L \times h}$ (one query per landmark) attend to the fused tokens
via multi-head cross-attention, yielding phase-aware embeddings
\begin{equation}
E^{\phi} = \mathrm{MHA}(Q^{\phi}, z, z) \in \mathbb{R}^{B \times L \times h},
\qquad \phi \in \{\mathrm{ED}, \mathrm{ES}\},
\end{equation}
with $h\!=\!32$ and $4$ heads. Each embedding is decoded into a chord by a
gated MLP and constrained to the image grid by a scaled $\tanh$:
\begin{equation}
c^{\phi} = \mathrm{GLU}\big(\mathrm{MLP}(E^{\phi})\big) \in
\mathbb{R}^{B \times L \times 4}, \qquad
\ell^{\phi} = \tfrac{\rho}{2}\big(\tanh(c^{\phi}) + 1\big),
\end{equation}
where each chord is parameterized by two endpoints $(x_1,y_1,x_2,y_2)$ and
$\rho\!=\!112$ is the image size. Stacking both phases gives the predicted
borders $\mathrm{EBs}=\{\ell^{\mathrm{ED}},\ell^{\mathrm{ES}}\} \in
\mathbb{R}^{B \times 2 \times L \times 4}$ and landmark embeddings
$\mathrm{LEs} \in \mathbb{R}^{B \times 2 \times L \times h}$. EBs provide
explicit border supervision and visualization, while LEs carry a phase-aware signal to the EF head.

\subsection{Endocardial Feature Aggregator (E\texorpdfstring{$^2$}{2}FA)}
EF estimation also benefits from global descriptors of LV contractility over
the cardiac cycle. E\textsuperscript{2}FA summarizes the deepest feature map
$x$ with three global statistics across the spatio-temporal dimensions, average, maximum, and variance concatenated into a global descriptor
$x_{\text{glob}} \in \mathbb{R}^{B \times 3D}$. In parallel, the landmark embeddings are flattened and projected to the
backbone width by $\mathrm{MLP}_{\mathrm{L}}$, giving a per-sample landmark
descriptor $x_{\text{lnd}} = \mathrm{MLP}_{\mathrm{L}}(\mathrm{flatten}
(\mathrm{LEs})) \in \mathbb{R}^{B \times D}$. Global and landmark descriptors
are fused and regressed into a bounded EF:
\begin{equation}
F_{\text{final}} = [\,x_{\text{glob}}, x_{\text{lnd}}\,], \qquad
\widehat{\mathrm{EF}} = \tfrac{1}{2}\big(\tanh(\mathrm{MLP}_{\text{head}}
(F_{\text{final}})) + 1\big)\times 100.
\end{equation}

\subsection{Landmark-Based Geometric Training Objective}
\label{sec:loss}
Simpson's method of disks estimates LV volume by dividing the chamber into a
stack of elliptical disks along the LV long axis;
EF then follows from end-diastolic and end-systolic volumes as
$\mathrm{EF}=(\mathrm{EDV}-\mathrm{ESV})/\mathrm{EDV}$. We translate this
geometry into a differentiable regularizer over the predicted chords.
The primary loss term is the mean-squared EF regression loss,
$\mathcal{L}_{\text{EF}} = \frac{1}{N}\sum_n(\widehat{\mathrm{EF}}_n -
\mathrm{EF}^{\text{gt}}_n)^2$. For the geometric surrogate, we treat the chord
$i$ of phase $\phi$ as a disk cross-section: its diameter is the chord
width $B_{\phi,i} = \lVert \mathbf{p}^{(1)}_{\phi,i} -
\mathbf{p}^{(2)}_{\phi,i}\rVert_2$, and its height $H_{\phi,i}$ is the
perpendicular distance from the center of chord $i\!-\!1$ to the line through
chord $i$. These yield a disk stack with per-disk volume $\pi
(B_{\phi,i}/2)^2 H_{\phi,i}$, recovering EDV/ESV. Three geometric terms supervise complementary aspects of this
construction:

\begin{equation}
\mathcal{L}_{\text{geo}} = \underbrace{\mathcal{L}_{\text{pts}}}_{\text{chord
localization}} + \underbrace{\mathcal{L}_{B}}_{\text{disk diameters}}
+ \underbrace{\mathcal{L}_{H}}_{\text{long-axis spacing}},
\end{equation}
where $\mathcal{L}_{\text{pts}}$ is the MSE between predicted and reference
chord endpoints, $\mathcal{L}_{B}$ and $\mathcal{L}_{H}$ match disk diameters
and heights. Each term captures a distinct geometric property
(landmark position, disk diameter, and long-axis spacing). The full objective is
\begin{equation}
\mathcal{L}_{\text{total}} = \mathcal{L}_{\text{EF}} + \lambda_{\text{geo}}\,
\mathcal{L}_{\text{geo}}, \qquad \lambda_{\text{geo}} = 0.05,
\end{equation}
with the geometric terms equally weighted under $\lambda_{\text{geo}}$.
\section{Experiments}
\label{sec:experiments}

\textbf{Datasets and Implementation.}
We evaluate on two public benchmarks. EchoNet-Dynamic~\cite{ouyang2020video}
comprises 10{,}030 apical four-chamber (A4C) echocardiography videos with
expert LVEF labels and ED/ES left-ventricular tracings; we use the official
training, validation, and test splits. EchoNet-Pediatric~\cite{reddy2023echonetpeds}
provides pediatric studies acquired in a parasternal short-axis (PSAX) view, where we follow an $80\%/10\%/10\%$ split, yielding 3443 training, 430 validation, and 431 held-out test videos. Videos are processed at $112\times112$ resolution. Following prior
work~\cite{muhtaseb2022echocotr}, we sample $64$ frames at a sampling period
of $2$; for shorter clips we pad with the running average of previous frames.
Echo-E\textsuperscript{3}Net is trained for $45$ epochs on a single NVIDIA RTX
4070 (12\,GB) with batch size $8$, using AdamW with learning rate $5\times10^{-4}$ and weight decay of $10^{-4}$. A key
practical advantage over prior work is that our model attains strong
performance without external pre-training, or heavy data augmentation; in contrast, EchoGNN~\cite{mokhtari2022echognn} relies on
ED/ES classification pre-training, EchoCoTr~\cite{muhtaseb2022echocotr} on
vision-domain pre-trained weights, and EchoGraphs~\cite{thomas2022light} on
extensive augmentation. We measure agreement with mean absolute error (MAE),
root mean squared error (RMSE), and R\textsuperscript{2}, and characterize
efficiency with parameter count and GFLOPs.

\subsection{Comparison with State of the Art}
Table~\ref{tab:ef_comparison} compares Echo-E\textsuperscript{3}Net against
recent EF regression baselines on EchoNet-Dynamic, each evaluated in its
original configuration. Echo-E\textsuperscript{3}Net achieves competitive accuracy (MAE $3.92$, RMSE $5.20$, R\textsuperscript{2}
$0.82$) while operating at a fraction of the computational cost. Specifically,
at just $8.05$ GFLOPs it reduces compute by $86.3\%$ relative to
CoReEcho~\cite{maani2024coreecho} ($58.92$G) and by over $99.9\%$ relative to
CardiacNet~\cite{yang2025cardiacnet} ($7949$G), and with only $1.55$M
parameters it is $92.7\%$ and $94.5\%$ more compact than these two methods,
respectively. We further assess the real-time behavior of Echo-E\textsuperscript{3}Net by measuring inference latency. At the EchoNet-Dynamic acquisition rate of
$\sim$50\,fps, an input clip spans roughly 2.56\,s of native acquisition,
covering two to three cardiac cycles. Echo-E\textsuperscript{3}Net processes a
clip in a median of 112.7\,ms on a CPU-only workstation (12 threads, batch
size 1), roughly $23\times$ faster than the acquisition itself. EF is
therefore available effectively as soon as the loop is recorded, without GPU
hardware, supporting point-of-care use.

% --- table ---
\begin{table*}[t]
\centering
\small
\caption{Comparison of EF estimation models on EchoNet-Dynamic. Parameter
count in millions. \colorbox{bestgreen}{Green} indicates the best result and
\colorbox{secondyellow}{yellow} the second best.}
\label{tab:ef_comparison}
\begin{adjustbox}{max width=\textwidth}
\begin{tabular}{|l|c|c|c|c|c|c|}
\hline
\textbf{Model} & \textbf{Frames} & \textbf{FLOPs} & \textbf{Params} &
\textbf{MAE}\,$\downarrow$ & \textbf{RMSE}\,$\downarrow$ & \textbf{R\textsuperscript{2}}\,$\uparrow$ \\
\hline
UVT~\cite{reynaud2021ultrasound}      & 128 & 130.00G & -    & 5.95 & 8.38 & 0.52 \\
MC3~\cite{ouyang2020video}            & 32  & 97.656G & -    & 4.45 & 5.68 & 0.76 \\
EchoGNN~\cite{mokhtari2022echognn}    & 64  & -       & \sbest{1.7}  & 4.45 & -    & 0.77 \\
EchoNet-Dynamic~\cite{ouyang2020video}& 32  & 91.974G & 32   & 4.22 & 5.56 & 0.79 \\
EchoGraphs~\cite{thomas2022light}     & 16  & 40.739G & 27.6 & 4.01 & 5.36 & \sbest{0.81} \\
EchoCoTr~\cite{muhtaseb2022echocotr}  & 36  & 44.907G & 21   & 3.98 & 5.34 & \sbest{0.81} \\
CoReEcho~\cite{maani2024coreecho}     & 36  & 58.92G  & 21   & \sbest{3.90} & \best{5.13} & \best{0.82} \\
CardiacNet~\cite{yang2025cardiacnet}  & 16  & 7949G   & 28   & \best{3.83}  & -    & -    \\
\textbf{Echo-E\textsuperscript{3}Net (Ours)} & 64 & \best{8.05G} & \best{1.55} & 3.92 & \sbest{5.20} & \best{0.82} \\
\hline
\end{tabular}
\end{adjustbox}
\end{table*}
\
To test the generalization of our method, we then extend our evaluation to the EchoNet-Pediatric dataset under matched settings
(Table~\ref{tab:pediatric}). Echo-E\textsuperscript{3}Net achieves the best
MAE ($3.88$) and R\textsuperscript{2} ($0.73$) while remaining the most compact
model, outperforming EchoCoTr~\cite{muhtaseb2022echocotr} and
CoReEcho~\cite{maani2024coreecho} despite using an order of magnitude fewer
parameters. Although the disk-based formulation is defined along the A4C long axis, the underlying supervision constrains ED and ES endocardial geometry. In PSAX videos, these constraints serve as a view-agnostic prior on LV shape and contraction, while the surrogate is not interpreted as exact disk integration. The consistent improvement on PSAX suggests that the anatomical signal remains useful beyond the apical view.

\subsection{Ablation Study \& Interpretability}
Table~\ref{tab:ablation} reports an ablation over the two modules
and the three geometric loss terms. Removing either module degrades all three
metrics: dropping E\textsuperscript{2}CBD (and, with it, the geometric
supervision) raises RMSE from $5.20$ to $5.40$, and dropping
E\textsuperscript{2}FA raises it to $5.31$, confirming that explicit endocardial
geometry and global contractility statistics are complementary. Each geometric
loss term yields a small but consistent gain: removing $\mathcal{L}_{\text{pts}}$,
$\mathcal{L}_{B}$, or $\mathcal{L}_{H}$ individually increases RMSE and reduces
R\textsuperscript{2} relative to the full model.
This indicates the terms are non-redundant, each constraining a distinct
geometric property (landmark position, disk diameter, and long-axis spacing),
and that the full Simpson-inspired objective is best.
%%%%%%%%%%%%%%%%%%%%%%%%%%%%%
\begin{table}[t]
\centering
\caption{Ablation over Echo-E\textsuperscript{3}Net components on
EchoNet-Dynamic. Results are averaged over 3 runs. \colorbox{bestgreen}{Green} best,
\colorbox{secondyellow}{yellow} second best.}
\label{tab:ablation}
\setlength{\tabcolsep}{5pt}
\renewcommand{\arraystretch}{1.2}
\footnotesize
\begin{tabular}{ccccccccc}
\toprule
\multicolumn{5}{c}{\textbf{Component}} & \multicolumn{3}{c}{\textbf{Metric}}\\
\cmidrule(lr){1-5}\cmidrule(lr){6-8}
E\textsuperscript{2}CBD & E\textsuperscript{2}FA &
$\mathcal{L}_{\text{pts}}$ & $\mathcal{L}_{B}$ & $\mathcal{L}_{H}$ &
MAE\,$\downarrow$ & RMSE\,$\downarrow$ & R\textsuperscript{2}\,$\uparrow$ \\
\midrule
\ding{52} & \ding{52} & \ding{56} & \ding{52} & \ding{52} & 3.98 & 5.30 & 0.81 \\
\ding{52} & \ding{52} & \ding{52} & \ding{56} & \ding{52} & 4.00 & 5.31 & 0.81 \\
\ding{52} & \ding{52} & \ding{52} & \ding{52} & \ding{56} & \sbest{3.95} & \sbest{5.29} & 0.81 \\
\ding{56} & \ding{52} & --        & --        & --        & 4.05 & 5.40 & 0.80 \\
\ding{52} & \ding{56} & \ding{52} & \ding{52} & \ding{52} & 4.01 & 5.31 & 0.81 \\
\rowcolor{bestgreen}
\ding{52} & \ding{52} & \ding{52} & \ding{52} & \ding{52} & 3.92 & 5.20 & 0.82 \\
\bottomrule
\end{tabular}
\end{table}
%%%%%%%%%%%%%%%%%%%%%%%%%%%%%

\begin{table}[t]
\centering
\caption{Generalization to EchoNet-Pediatric. \colorbox{bestgreen}{Green}
best, \colorbox{secondyellow}{yellow} second best.}
\label{tab:pediatric}
\setlength{\tabcolsep}{8pt}
\renewcommand{\arraystretch}{1.2}
\footnotesize
\begin{tabular}{lccc}
\toprule
Model & Params & MAE\,$\downarrow$ & R\textsuperscript{2}\,$\uparrow$ \\
\midrule
EchoGNN~\cite{mokhtari2022echognn}   & \sbest{1.7\,M} & 5.18 & 0.44 \\
EchoCoTr~\cite{muhtaseb2022echocotr} & 21\,M & \sbest{4.13} & \sbest{0.70} \\
CoReEcho~\cite{maani2024coreecho}    & 21\,M & 4.44 & 0.67 \\
\textbf{Echo-E\textsuperscript{3}Net} & \best{1.55\,M} & \best{3.88} & \best{0.73} \\
\bottomrule
\end{tabular}
\end{table}

\begin{figure*}[h]
    \centering
    \includegraphics[width=\textwidth]{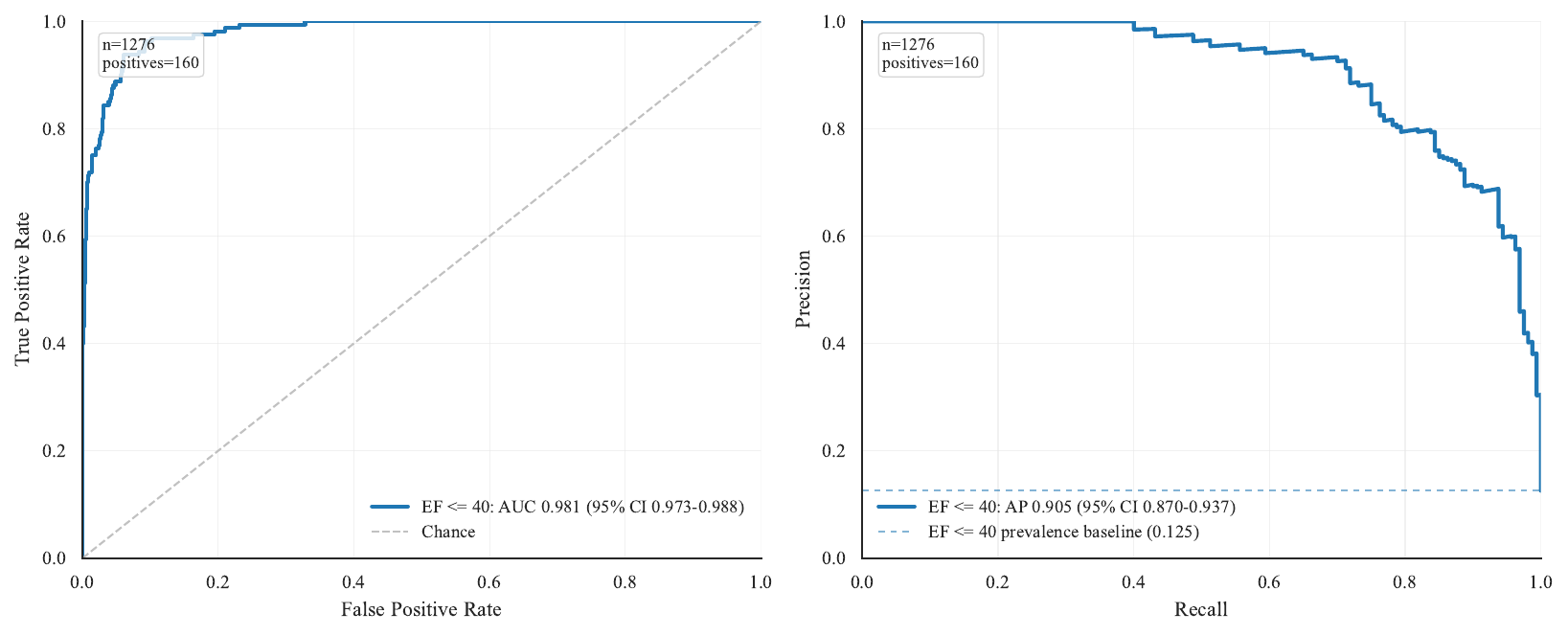}
    \caption{Discrimination for reduced ejection fraction on the EchoNet-Dynamic test set using a
threshold of LVEF $\leq$ 40. Left, receiver operating characteristic curve
with area under the curve and confidence interval. Right, precision recall
curve with average precision and confidence interval; the dashed line
indicates the prevalence baseline.}
    \label{fig:roc_pr_combined}
\end{figure*}

\begin{figure*}[h]
    \centering
    \includegraphics[width=\textwidth]{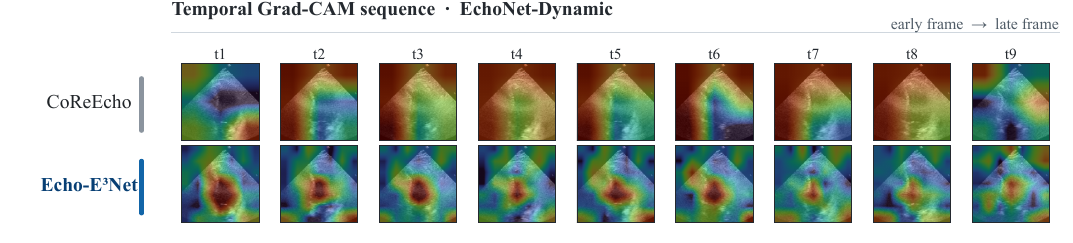}
    \caption{Grad-CAM~\cite{selvaraju2017grad} across the cardiac cycle on an
EchoNet-Dynamic video. Echo-E\textsuperscript{3}Net's saliency tracks the
contracting left-ventricular cavity frame-to-frame, whereas CoReEcho's drifts
into surrounding tissue, an effect of E\textsuperscript{2}CBD's explicit
endocardial supervision.}
    \label{fig:attention}
\end{figure*}

\Cref{fig:attention} shows a temporal Grad-CAM~\cite{selvaraju2017grad}
sequence across the cardiac cycle for Echo-E\textsuperscript{3}Net and
CoReEcho~\cite{maani2024coreecho}. Our model's
activations remain tightly concentrated on the left-ventricular cavity and
endocardial border across frames, with limited response in surrounding tissue,
whereas the baseline's are more diffuse.
We further assess clinical utility for detecting heart failure with reduced
ejection fraction (LVEF $\leq 40$) on the EchoNet-Dynamic test set ($n{=}1276$,
$160$ positives). Using the continuous predicted LVEF as a decision score,
Echo-E\textsuperscript{3}Net attains an area under the ROC curve of $0.981$
($95\%$ CI $0.973$--$0.988$) and an average precision of $0.905$ ($95\%$ CI
$0.870$--$0.937$), far exceeding the prevalence baseline of $0.125$
(\Cref{fig:roc_pr_combined}).
%%%%%%%%%%%%%%%%%%%%%%%%%%%%%

%%%%%%%%%%%%%%%%%%%%%%%%%%%%%%%%
\section{Discussion and Limitations}
\label{sec:discussion}
Our proposed Echo-E\textsuperscript{3}Net, with explicit anatomical guidance
embedded into LVEF estimation, achieves competitive accuracy at a fraction of
the compute of existing methods, generalizes across views, and remains
deployable on CPU hardware.
\\
Several limitations remain. First, on EchoNet-Dynamic, our accuracy is on par
with the strongest baselines; we view the present
contribution as establishing a strong accuracy--efficiency operating point, and
leave to future work the question of pushing accuracy higher while preserving
the lightweight design. Second, although we evaluate on two views and two
populations, generalization across scanners, vendors, and acquisition
protocols remains to be established on external multi-centre cohorts. Third,
the predicted endocardial chords serve only as a training-time geometric
regularizer; producing clinically faithful, orderable contours suitable for
segmentation would require explicit ordering supervision, which we leave to
future work.

\section{Conclusion}
\label{sec:conclusion}
We introduced Echo-E\textsuperscript{3}Net, an efficient, anatomy-guided
spatio-temporal network for LVEF estimation. A dual-phase endocardial border
detector and an endocardial feature aggregator, trained with a Simpson-inspired
geometric objective, deliver accuracy on par with the state of the art on
EchoNet-Dynamic and the best results on EchoNet-Pediatric, using only $1.55$M
parameters and $8.05$ GFLOPs with real-time CPU inference. These properties
make Echo-E\textsuperscript{3}Net well suited to resource-constrained
point-of-care ultrasound. 

\subsubsection{Acknowledgments.} This work was supported by the Canadian Foundation for Innovation-John R. Evans Leaders Fund (CFI-JELF) program grant number 42816. Mitacs Accelerate program grant number AWD024298-IT33280. We also acknowledge the support of the Natural Sciences and Engineering Research Council of Canada (NSERC), [RGPIN-2023-03575]. Cette recherche a été financée par le Conseil de recherches en sciences naturelles et en génie du Canada (CRSNG), [RGPIN-2023-03575].  

\bibliographystyle{splncs04}
\bibliography{ref}

\end{document}